# Properties of Decameter IIIb–III Pairs

V.N. Melnik (1), A.I. Brazhenko (2), A.V. Frantsuzenko (2), V.V. Dorovskyy (1), H.O. Rucker (3)

1 - Institute of Radio Astronomy, National Academy of Sciences of Ukraine, Chervonopraporna St., 4, Kharkiv, 61002, Ukraine

2 - Poltava Gravimetrical Observatory, S. Subbotin Institute of Geophysics, National Academy of Sciences of Ukraine, Myasoiedova St. 27/29, Poltava, 36014, Ukraine

3 - Commission for Astronomy, Schmiedlstrasse 6, Graz, 8042, Austria

**Abstract**

A large number of Type IIIb–III pairs, in which the first component is a Type IIIb burst and the second one is a Type III burst, are often recorded during decameter Type III burst storms. From the beginning of their observation, the question of whether the components of these pairs are the first and the second harmonics of radio emission or not has remained open. We discuss properties of decameter IIIb–III pairs in detail to answer this question. The components of these pairs, Type IIIb bursts and Type III bursts, have essentially different durations and polarizations. At the same time their frequency drift rates are rather close, provided that the drift rates of Type IIIb bursts are a little larger those of Type III bursts at the same frequency. Frequency ratios of the bursts at the same moment are close to two. This points at a harmonic connection of components in IIIb–III pairs. At the same time there was a serious difficulty, namely, why the first harmonic had fine frequency structure in the form of striae and the second harmonic did not have it. Recently Loi, Cairns, and Li (*Astrophys.* **790**, 67, 2014) succeeded in solving this problem.The physical aspects of observational properties of decameter IIIb–III pairs are discussed and pros and cons of harmonic character of Type IIIb bursts and Type III bursts in IIIb–III pairs are presented. We conclude that practically all properties of IIIb–III pair components can be understood in the framework of the harmonic relation of components of IIIb–III pairs.

## 1. Introduction

The existence of harmonic III–III pairs (so-called fundamental–harmonic pairs) was established many years ago (Wild, Murray, and Rowe, 1954). The frequency ratio of harmonic to fundamental averages 1.8:1, with a range from 1.6:1 to 2:1 (Suzuki and Dulk, 1985). Fundamentals have high polarizations up to $70-80\%$ with an average value of $35\%$ (Dulk, Suzuki, and Sheridan, 1984). The polarization of the harmonic is generally smaller, on average only 11 %. The durations of fundamentals are some times smaller than the durations of the harmonic. It is six times smaller than for the harmonic at 38 MHz (Dulk, Suzuki, and Sheridan, 1984). Drift rates of decametre fundamentals and harmonics are mainly close and equal to $2-4$ MHz $s^{-1}$ (Melnik *et al.*, 2011). Heights and apparent source sizes of fundamentals and harmonics (both components are Type III bursts) are practically the same and both increase with wavelengths (Dulk, Suzuki, and Sheridan, 1984). In such a way the diameter of radiated regions are $6^{1}$, $11^{1}$, and $20^{1}$ at 169 MHz, 80 MHz, and 43 MHz, respectively (Dulk, Suzuki, and Sheridan, 1984). Radio fluxes of Type III bursts and their brightness temperatures increase with wavelength and vary over in the wide ranges from 10 s.f.u. to $10^4$ s.f.u. for fluxes and from $10^7$ K to $10^{12}$ K for brightness temperature in the meter band (Dulk, Suzuki, and Sheridan, 1984). In the case of Type IIIb bursts their visible sizes do not differ essentially from those of Type III bursts (Abranin *et al.*, 1976, 1978) and because fluxes of Type IIIb bursts are not smaller than fluxes of Type III bursts their brightness temperatures are not smaller than those of Type III bursts. Taking into account scattering effects, that are important in the location of generation of the fundamental, leads to increase of brightness tempersture of Type IIIb bursts. Only in 10 % of all harmonic pairs is the fundamental Type IIIb burst (Dulk, Suzuki, and Sheridan, 1984). In these cases IIIb–III pairs have the same properties as III–III pairs except for the fine frequency structure of Type IIIb bursts in the form of stria-bursts (Abranin *et al.*, 1976, 1978, 1979). Decame- ter IIIb–III pairs observed in July – August 2002 were discussed by Melnik *et al.* (2011). Their main properties were the same as for fundamental–harmonic pairs. The frequency ratio was closed to two. The frequency drift rates of Type IIIb bursts were larger than those for type III bursts in IIIb–III pairs and durations of Type IIIb bursts were four to five times smaller than Type III burst durations. According to Abranin *et.al* (1976, 1979) the sizes and locations of Type IIIb bursts and Type III bursts were practically the same. So the average sizes of sources for both Type IIIb and Type III bursts were equal to $30'$ at 25 MHz. Heights of sources of type IIIb and type III bursts in the solar corona have been in the range from 1.75 Rs to 2.22 Rs at 25 and 12.5 MHz, respectively (Abranin *et al.*, 1976).

Contrary to the harmonic character of IIIb–III pairs (Smerd, 1976) there was the hypothesis that Type IIIb bursts were precursors of Type III bursts, which were generated by the electrons with larger velocities than those of electrons that are responsible for the following Type III burst (de La Noe and Boischot, 1972, Benz *et al.*, 1982). It follows that the drift rates of Type IIIb bursts are larger than the Type III drift rates. This hypothesis is supported by the facts that the heights and sizes of Type IIIb bursts and Type III bursts are practically the same.

From the beginning, one of the serious difficulties of the IIIb–III harmonic character was the fine frequency structure of Type IIIb bursts and the absence of such structure at Type III bursts (Abranin *et al.*, 1979, 1984). The fine frequency structure of Type IIIb bursts manifests itself as a serious problem. As early as in the first article (Takakura and Yousef, 1975), in which Type IIIb bursts have been considered, Takakura proposed that electron beams propagating through coronal plasma with inhomogeneity generated radio emission in the form of inhomogeneously structured striae. This idea was developed by articles (Kontar, 2001, Kontar and Pécseli, 2002, Loi, Cairns, and Li, 2014). It was numerically shown by Kontar (2001) that there were regions where the level of Langmuir waves generated by electron beams was higher and lower depending on variations of coronal density through which these electrons are passing. As a result, radio emissions from different regions will be of different intensities and in principle some fine frequency structure can be obtained. Returning to the harmonic character of IIIb–III pair there is the question of why, if these pairs are harmonic, a Type III burst does not have such a fine frequency structure? Recently Loi, Cairns, and Li (2014) have numerically shown that when propagating through a plasma with Kolmogorov turbulence an electron beam can generate fundamental radio emission in the form of Type IIIb bursts and the second harmonic with a smooth profile like Type III bursts. However the durations of both the fundamental and harmonic appeared to be practically the same. It should be said that the differences both in durations and in frequency drift rates of Type IIIb bursts and Type III bursts have not practically been discussed at all.

The problem concerning the identical sizes and the same heights of Type IIIb and Type III bursts can be resolved in the model of ducts (Duncan, 1979). Also an interesting opportunity was proposed by Eremin and Zaitsev (1985). They considered the situation that the fundamental with frequency $f$ propagating through the solar corona split into two Langmuir waves of half frequencies $f/2$ at the location where the local plasma frequency $f_{pe} = f/2$ and after that they coalesced again in the electromagnetic wave with the frequency $f$. At this location the harmonic with frequency $f$ is also generated when fast electrons arrive. As a result the secondary fundamental and the harmonic emerge from the same place.

Therefore there is evidence in favour of the harmonic character of IIIb–III pairs and also against it.

In this article observational properties of IIIb–III pairs observed by the radio telescope URAN-2 in April, June, and September 2011 are considered. Drift rate, duration, and polarization dependences on frequency are analyzed to clarify the harmonic character of components of IIIb–III pairs. Interpretations of these properties are discussed.

## 2. Observations

IIIb–III pairs were observed by the radio telescope URAN-2 (Poltava, Ukraine) in April, June, and September 2011. The radio telescope consists of 512 cross-dipoles situated at 45° to meridian and has an area of 28,000 m$^2$ with an antenna beam of size $3° \times 7°$ (Megn *et al.*, 2003; Brazhenko *et al.*, 2005) . Observations were made with the DSPZ *(Digital wideband frequency Spectral Polarimeter)* at frequencies 8 – 32 MHz in April and September 2011 and at frequencies 16 – 32 MHz in June 2011 with a frequency–time resolution of 4 kHz – 100 ms (Ryabov *et al.*, 2010).

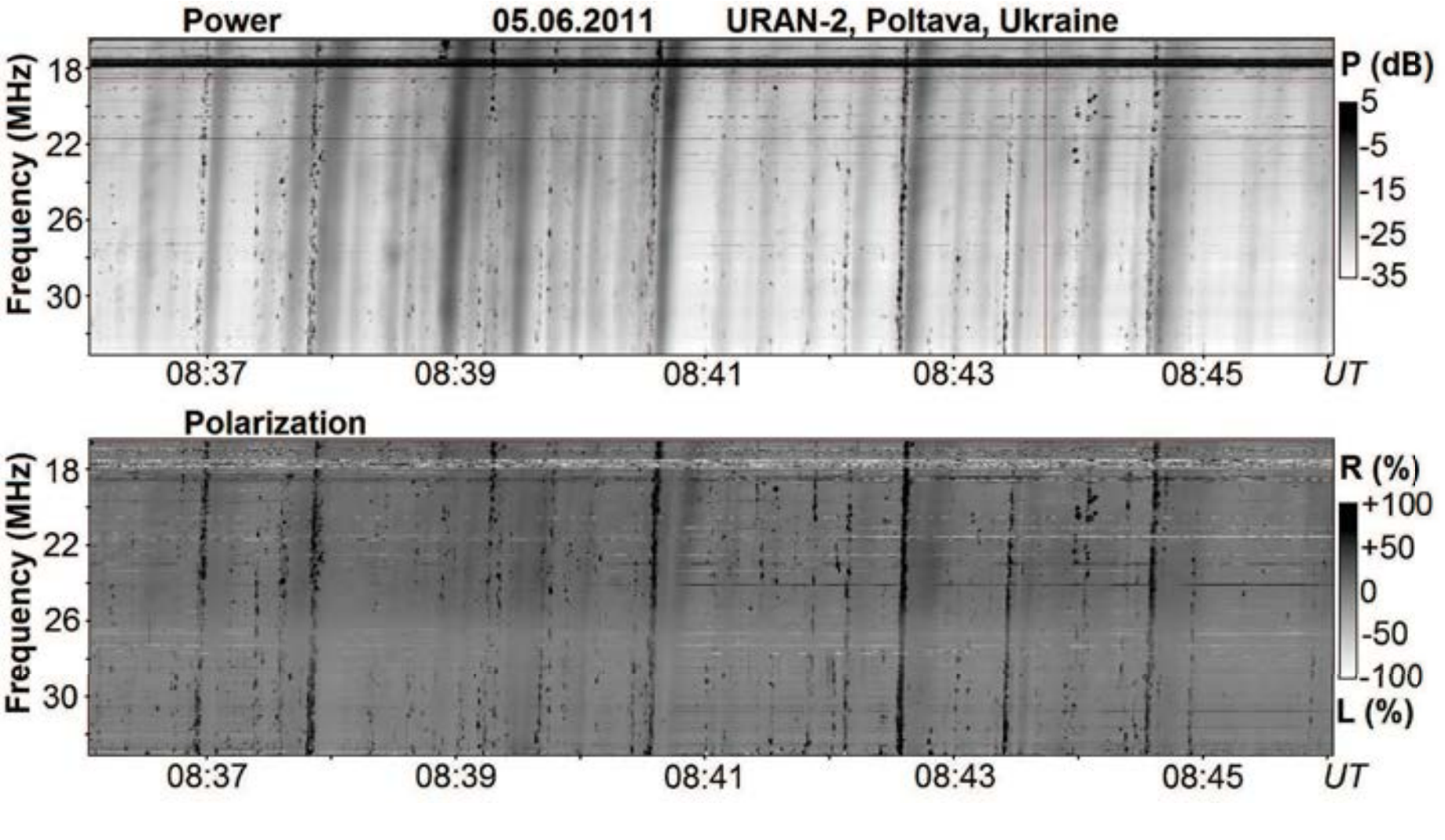

Figure 1. Fragment of Type IIIb bursts–Type III bursts storm with plenty of IIIb–III pairs on 5 June 2011.

We have chosen for an analysis three periods of observations: 1 – 7 April, 3 – 6

June and 4 – 7 September 2011 (Brazhenko *et al.*, 2015). In this time we registered storms of Type IIIb bursts and Type III bursts and a great number of IIIb–III pairs (Figure 1). On these days there were several active regions and we could not associate enhanced decameter radio activities with one of them in particular. Observations were mainly made from 5:30 to 14:30 UT. 143, 106, and 56 IIIb-III pairs were analyzed in April, June, and September respectively. Dynamic spectra of some IIIb–III pairs of different periods are shown in Figure 2. As can seen, stria-bursts are located irregularly along Type IIIb bursts. There are rather noticeable gaps between them, and sometimes stria-bursts are closely grouped. Two successive IIIb–III pairs have different density of stria (Figure 2a). Two Type III bursts in these pairs are smooth, but at frequencies where type IIIb bursts have more stria, they have greater fluxes. If components are harmonically then their radio emissions at the same frequency come from different heights and such correlation should not exit. Such correlation can be expected if the first component is caused by the fastest electrons of the electron beam responsible for the second component, as in the model of de La Noe, and Boishot (1972), and Benz *et al.* (1982).

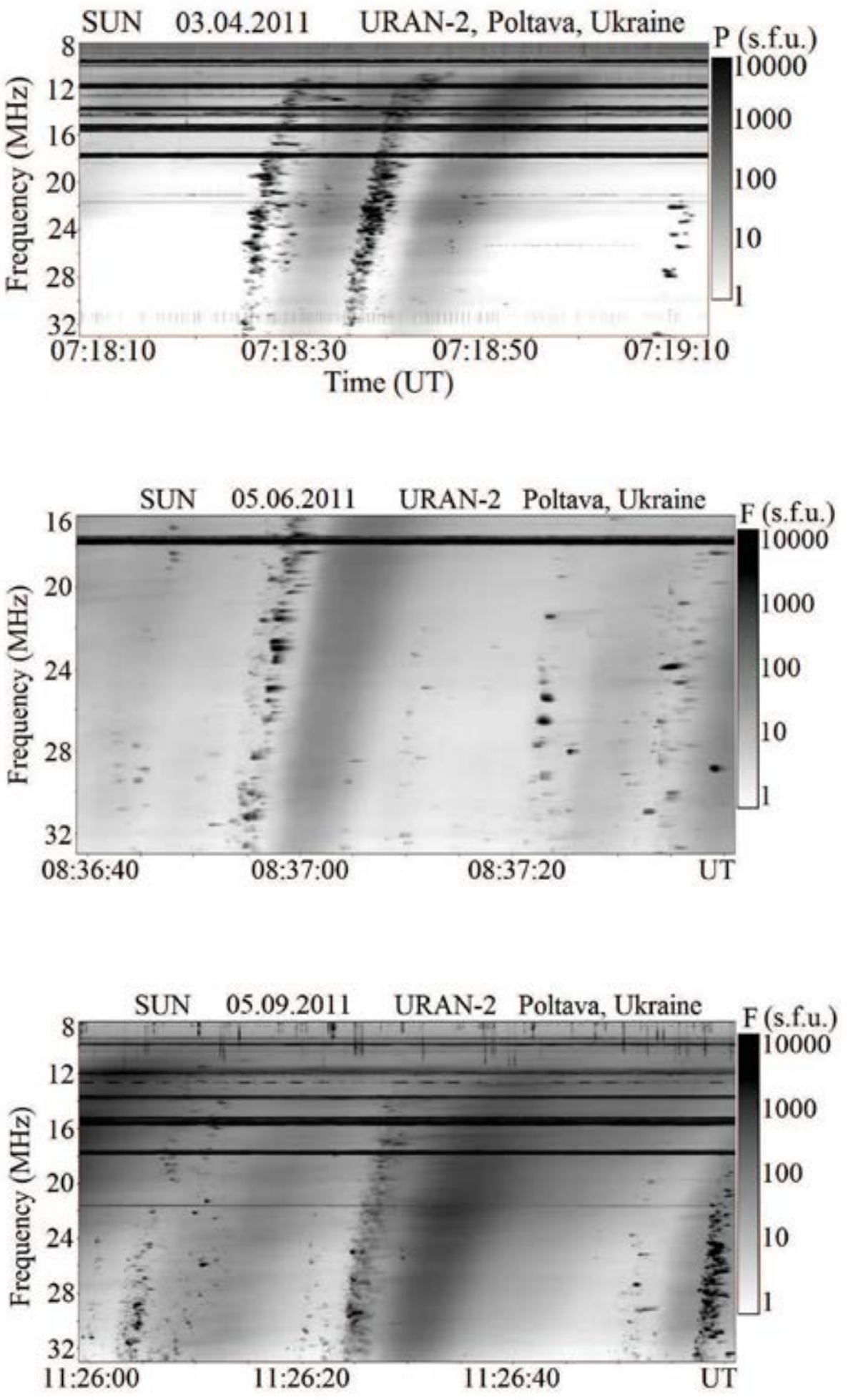

a)

b)

c)

Figure 2. Dynamic spectra of IIIb–III pairs observed on 3 April (a), 5 June (b), and 5 September (c) 2011.

As seen in Figure 3, Type IIIb bursts have larger radio fluxes than Type III bursts in the IIIb–III pairs. For 3 April and 5 June the average fluxes of Type IIIb are four to five times larger than those for Type III bursts in the whole frequency band from 10 to 32 MHz.

For 5 September the fluxes of both Type IIIb bursts and Type III bursts are practically equal at all frequencies. The flux– frequency dependences follow the power-law $I \propto f^{-\beta}$ with power $\beta \approx$ 0.8, 0.6, 1.0 for 3 April, 5 June, and 5 September, respectively. Maximum fluxes are a little more than 103 s.f.u. and minimum fluxes are about 10 s.f.u.

Profiles of some IIIb–III pairs observed on 3 April, 5 June, and 5 September are shown in Figure 4. It can be seen that type IIIb bursts consist of one, two, and even three stria-bursts at some frequencies. At the same time there are frequency bands without any stria (Figure 5). So, *e.g.*, the Type III burst observed on 5 June (Figure 5) has no precursor in the form of stria-burst of Type IIIb at frequency 27.3 MHz.

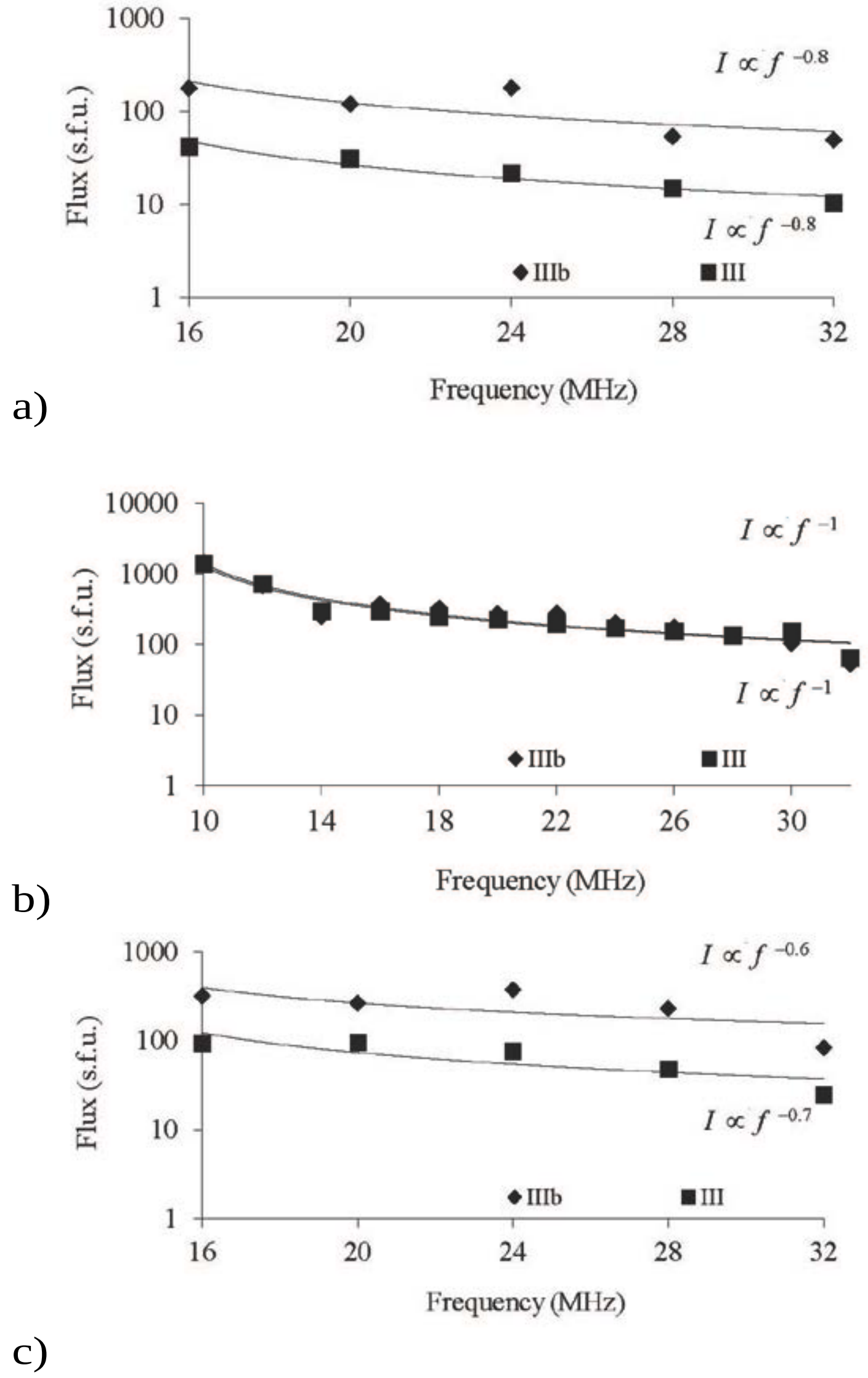

Figure 3. Flux dependences of Type IIIb and Type III bursts versus frequency for IIIb–III pairs on 3 April (a), 5 June (b) and 5 September (c) 2011.

Polarizations of Type IIIb bursts exceed significantly those of Type III bursts (Figure 4). Polarizations of Type III bursts are practically constant over the duration of the bursts. At the same time, polarizations of Type IIIb bursts achieve their maxima very quickly after the beginning of the bursts. After that the polarizations decrease regularly. So any polarization maximum is always preceding the flux maximum for all Type IIIb bursts (Figure 6).

We could not find any dependence between fluxes and polarizations for both Type IIIb bursts and Type III bursts in IIIb–III pairs. It was stated by Benz and Zlobec (1978) that there is a connection between frequency drift rates and polarizations of bursts. We have not found any such connections at frequencies 8 – 32 MHz (Figure 7).

We measured frequency drift rates, durations (half-maximum), polarizations and fluxes of bursts in frequency bands 10 – 14 MHz, 14 – 18 MHz, 18 – 22 MHz, 22 – 26 MHz, 26 – 30 MHz, and 30 – 33 MHz. Averaged drift rates, durations, and polarizations for Type IIIb bursts and Type III bursts at 16 MHz on 3 April, 5 June, and 5 September are presented in Table 1.

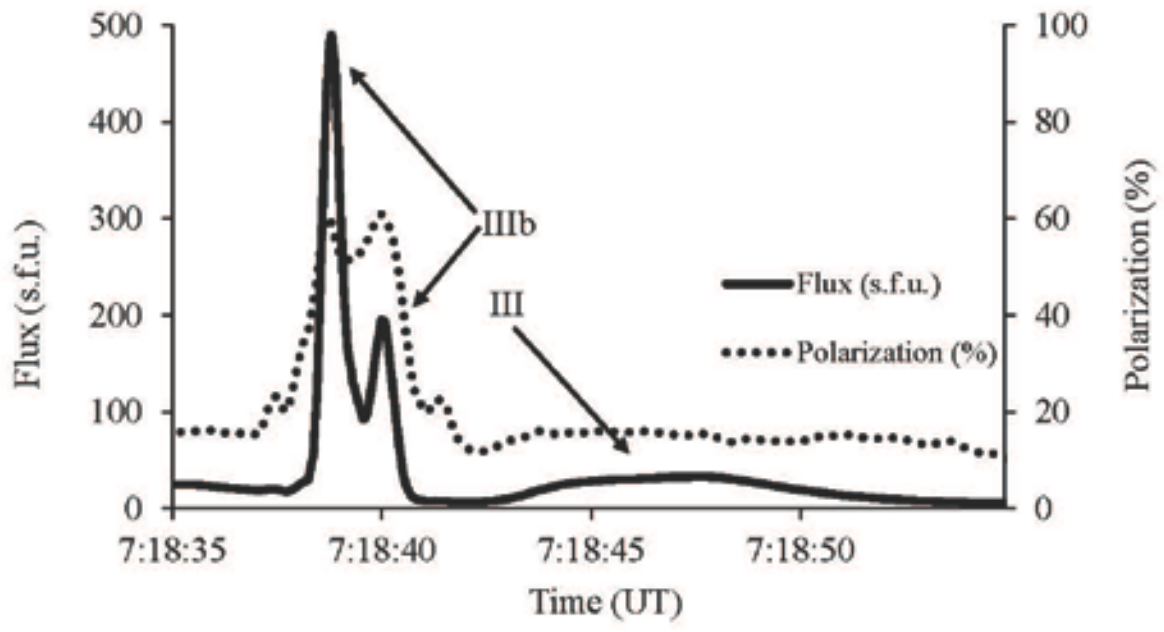

a)

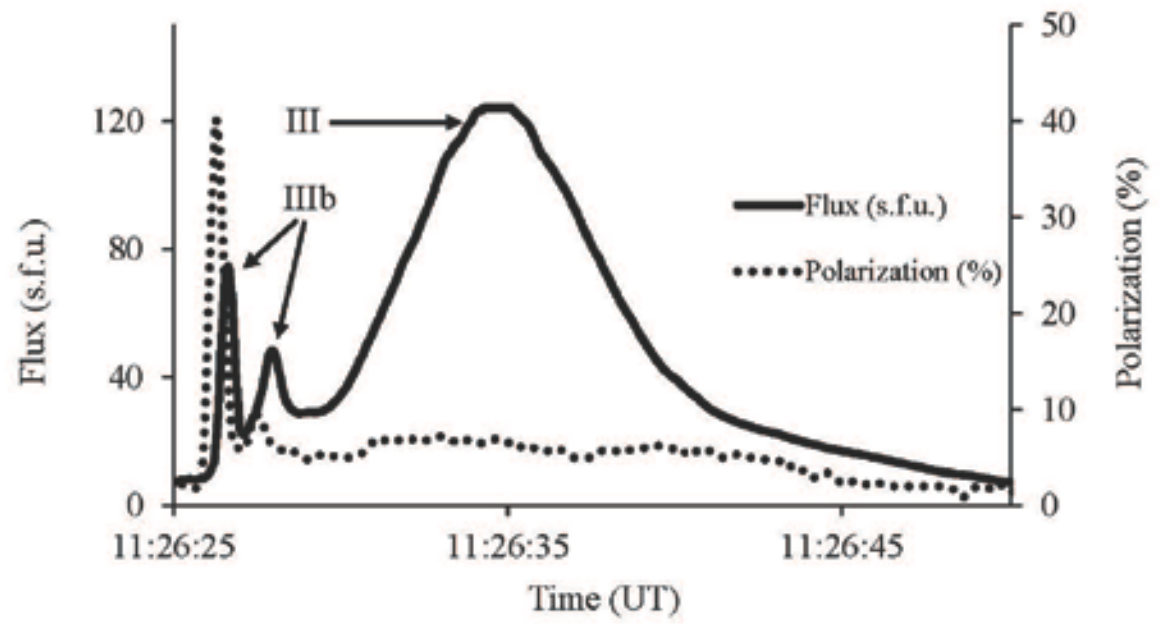

b)

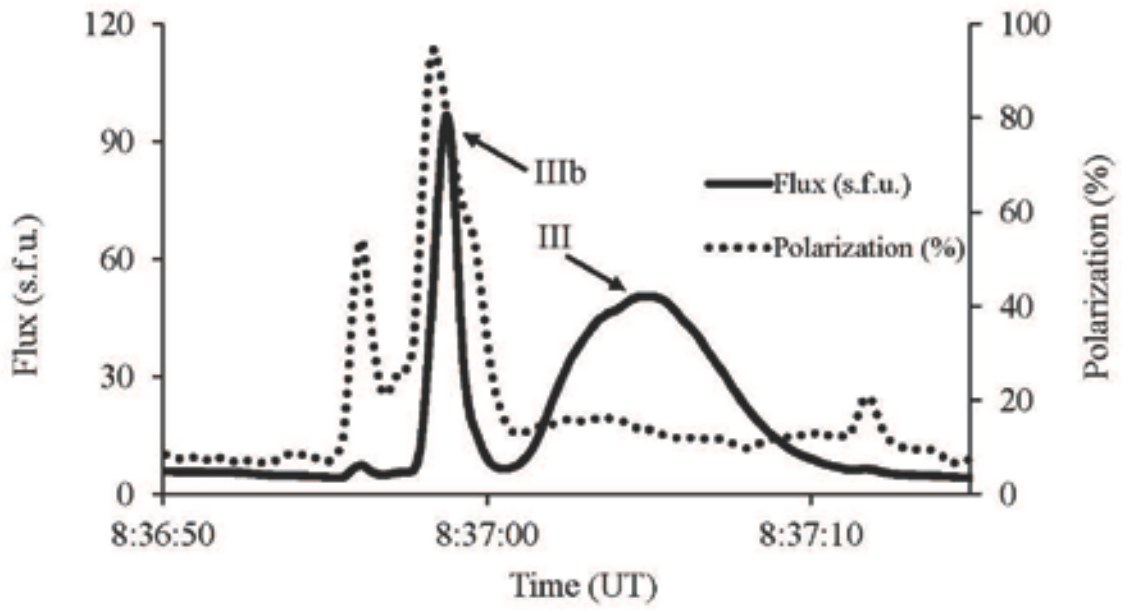

c)

Figure 4. Flux (solid curve) and polarization (dotted curve) profiles of IIIb–III pairs on 3 April (a), 5 June (b), and 5 September (c) 2011 at frequency 23 MHz.

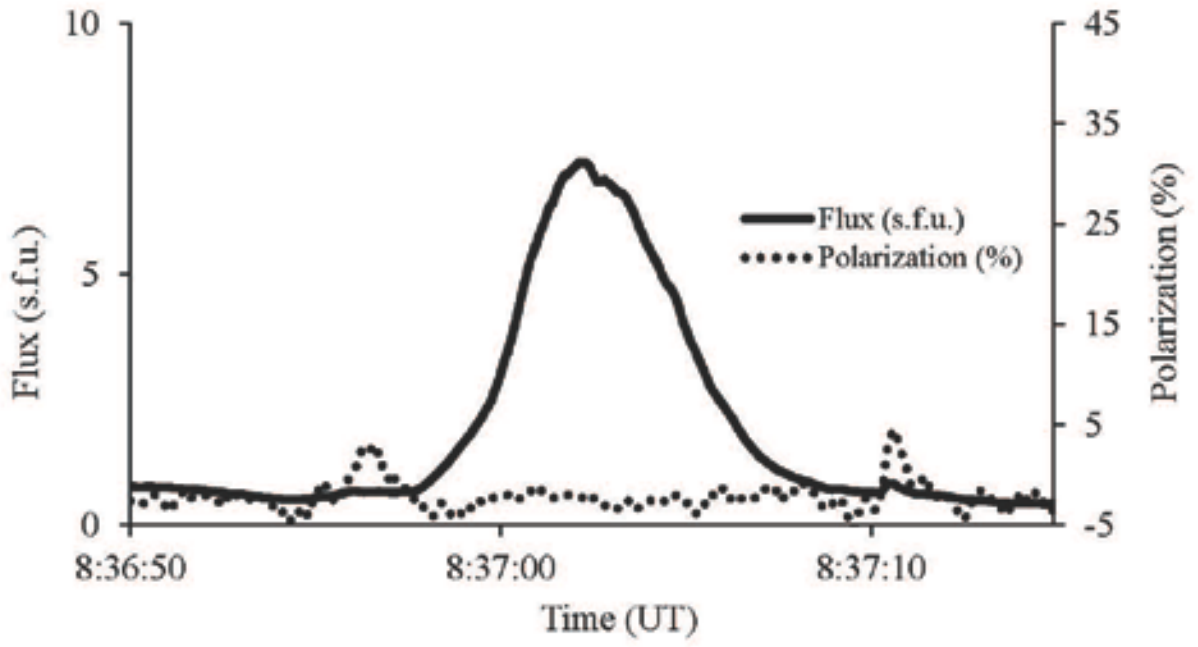

Figure 5. Profile of Type III burst observed on 5 June 2011 at frequency 27.3 MHz. Preceeding stria-bursts are absent.

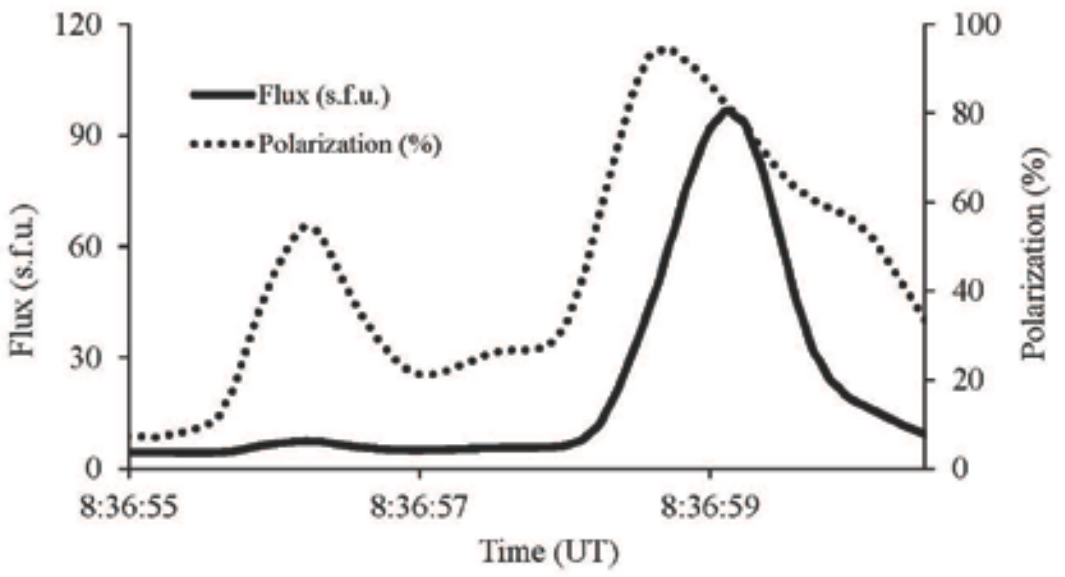

Figure 6. Flux (solid curve) and polarization (dotted curve) profiles for a Type IIIb burst observed on 5 June 2011 at frequency 23 MHz

We see that these values slightly change from month to month. Drift rates of Type IIIb bursts and Type III bursts are in the limits 2.2 – 4 MHz $s^{-1}$ and 1.5 – 4 MHz $s^{-1}$, respectively, provided that Type IIIb burst rates are always larger than Type III burst rates. Durations of Type IIIb bursts are essentially (4 – 7 times) smaller than those for Type III bursts. However, the polarizations of the former are larger (3 – 6 times) compared with polarizations of Type III bursts.

Figures 8 – 10 show the frequency dependences of averaged drift rates, dura- tions, and polarizations of Type IIIb bursts and Type III bursts for days with most IIIb–III pairs.

Drift–rate dependences for both components of IIIb–III pairs can be approximated by linear dependence

$$\frac{df}{dt} \approx -Af \tag{1}$$

Table 1. Frequency drift rates (one day averages), durations, and polarizations of Type IIIb bursts and Type III bursts at 16 MHz for 3 April, 5 June, and 5 September 2011.

| Parameters | Type IIIb bursts | | | Type III bursts | | |
|---|---|---|---|---|---|---|
| | 3 April | 5 June | 5 September | 3 April | 5 June | 5 September |
| Drift rate [ $MHz / s$ ] | 2.5 | 4.0 | 2.2 | 1.9 | 2.4 | 1.5 |
| Duration [s] | 2 | 1.2 | 1.9 | 11.7 | 8.6 | 9.4 |
| Polarization [%] | 50.3 | 59.6 | 31.8 | 11.4 | 11.1 | 12.2 |

Coefficients $A$ are presented in Table 2. They are close enough for Type IIIb bursts and Type III bursts for one and the very same day although they differ for different days. Durations of Type IIIb bursts and Type III bursts are decreasing with increasing frequency. Durations of Type IIIb bursts changed from 1 second at 16 MHz to 0.8 seconds at 32 MHz for 3 April and 5 June and

Table 2. Coefficient A for components IIIb–III pairs for different days of observations.

| Data | A [$s^{-1}$] | |
|---|---|---|
| | Type IIIb bursts | Type III bursts |
| 03 April 2011 | 0.17 | 0.18 |
| 05 June 2011 | 0.16 | 0.18 |
| 05 September 2011 | 0.08 | 0.11 |

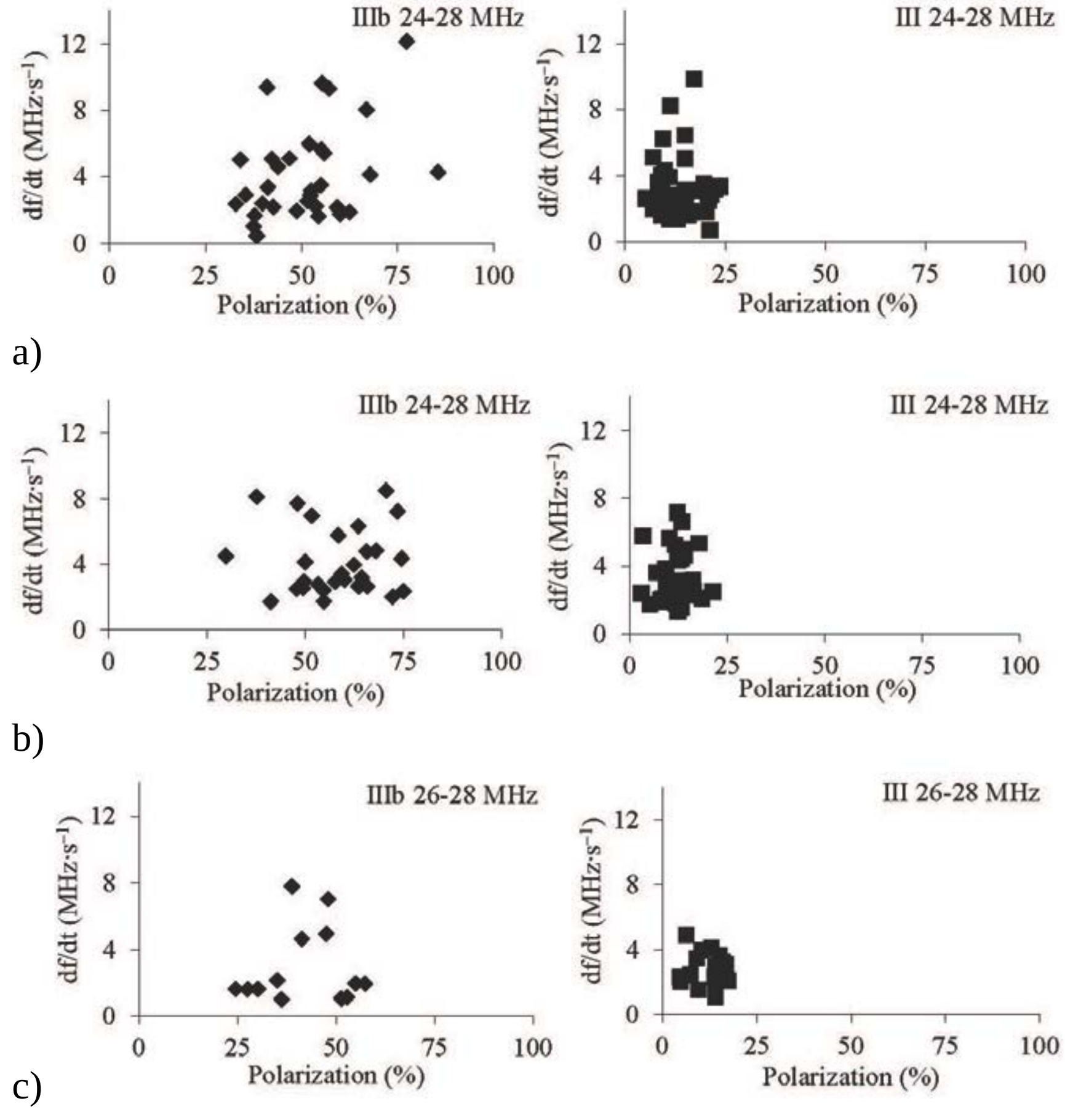

Figure 7. Frequency drift rates and polarizations for Type IIIb bursts and Type III bursts at different frequencies on 3 April (a), 5 June (b) and 5 September (c) 2011. Drift rates errors are about 20 % and polarizations errors are 10 – 15 % for Type IIIb bursts and 20 – 30 % for Type III bursts.

from 1.6 seconds at 16 MHz to 1.2 seconds at 32 MHz for 5 September. Type III bursts in IIIb–III pairs as well as usual Type III bursts have durations about 12 seconds at 16 MHz and only 5 seconds at 32 MHz (Mel'nik *et al.*, 2005). Note that durations of Type

IIIb bursts are essentially smaller than durations of Type III bursts. It is very difficult to explain.

Table 3. Power exponent [a] for Type IIIb bursts and Type III bursts in IIIb-III pairs for different days of observations.

—

| Data | a | |
|---|---|---|
| | Type IIIb bursts | Type III bursts |
| 03 April 2011 | $0.9 \pm 0.3$ | $0.5 \pm 0.2$ |
| 05 June 2011 | $0.7 \pm 0.2$ | $1.1 \pm 0.4$ |
| 05 September 2011 | $0.3 \pm 0.1$ | $1.0 \pm 0.3$ |

such a difference in the framework of harmonic relations of these bursts in IIIb–III pairs. Really, if both bursts are generated in the same place then their durations should be similar. Moreover, frequency dependences for durations for Type IIIb bursts and Type III bursts must also be similar. However, the situation is different. Usually the frequency dependences of burst durations are approximated by a power law (Elgaroy and Lyngstad, 1972, Suzuki and Dulk, 1985, Melnik *et al.*, 2011, Rutkevych and Melnik, 2012)

$$\tau \approx f^{-a} \qquad (2)$$

Power exponents [$a$] for Type IIIb bursts and Type III bursts are summarized in Table 3.

Comparing exponents [$a$] for Type IIIb bursts and Type III bursts we see that they vary notably for components in IIIb–III pairs. So we can conclude that durations of components in pairs depend probably on different processes.

Figure 10 shows that polarizations of Type IIIb bursts generally exceed those of Type III bursts. Polarizations of Type IIIb bursts were practically the same in April and June and had values of 60 % and even higher. In September the polarization was about 40 % on average. The polarizations of Type III bursts were within the range 10 – 15 %. There are no frequency dependencies for the polarizations of either Type IIIb bursts or Type III bursts. Such polarization properties of IIIb-III components support their harmonic character.

## 3. Discussion

In this section we shall discuss observational properties of IIIb–III pair components in detail.

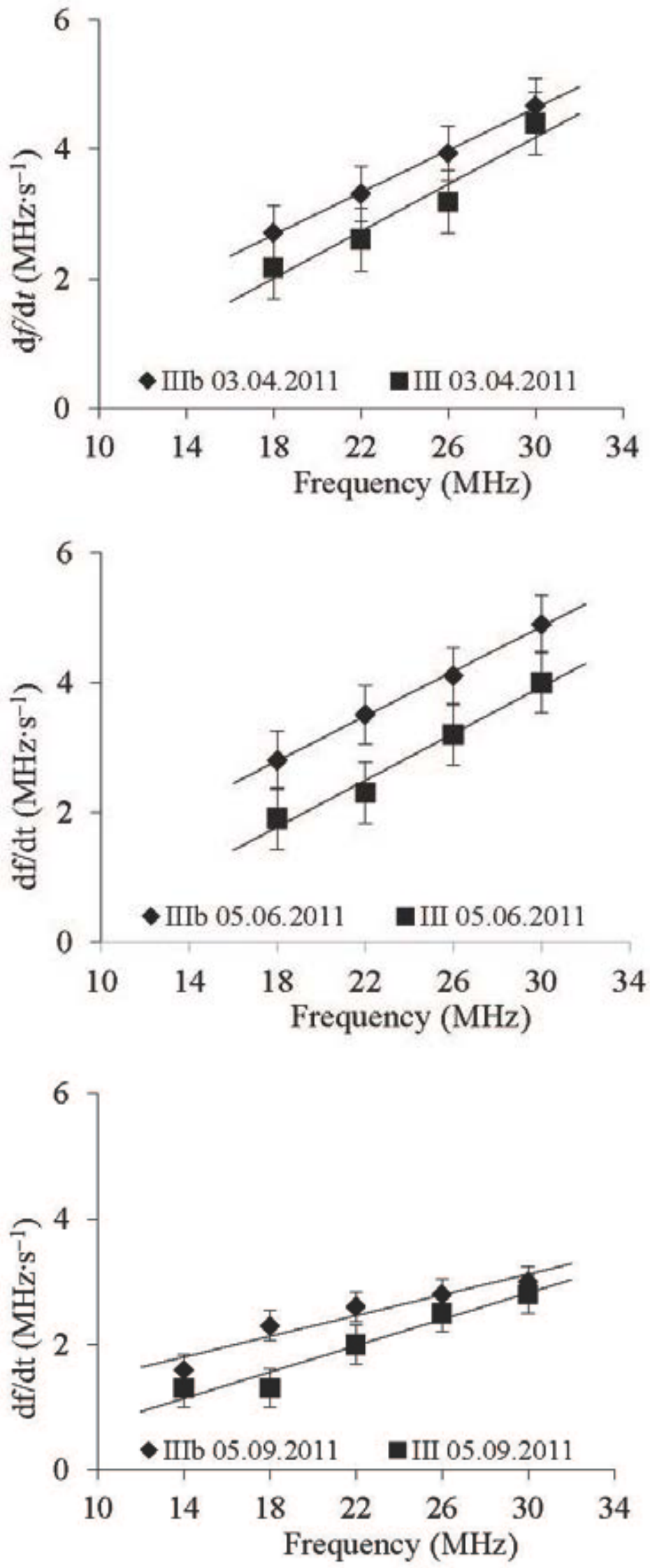

Figure 8. Drift-rate dependences of Type IIIb bursts and Type III bursts on frequency for 3 April (a), 5 June (b), and 5 September (c) 2011.

### 3.1 Frequency Ratio

Averaged frequency ratios for Type III bursts and Type IIIb bursts at one and the same time were derived for the days of maximum IIIb–III numbers in April, June, and September 2011. They are equal to 1.94±0.07, 1.93±0.04, and 1.87±0.09 for 3 April (44 pairs), 5 June (41 pairs), and 5 September (23 pairs), respectively. In the case where Type IIIb bursts are fundamentals and Type III bursts are harmonics, the frequency ratio should be 2:1. Ratios with a small difference from this value can be understood if we take into account the fact that the group velocities of fundamentals and harmonics are different. Indeed, in the location of the fundamental generation, its frequency is close to the local plasma frequency and as a consequence its group velocity is small. At the same time, because the harmonic has a frequency equal to twice the local plasma frequency, the harmonic group velocity is generally larger than that for the fundamental. As a result harmonics arrive at Earth earlier than fundamentals and thus the frequency ratio should be smaller than 2:1. So we conclude that the frequency ratio supports the harmonic connection between components of IIIb–III pairs.

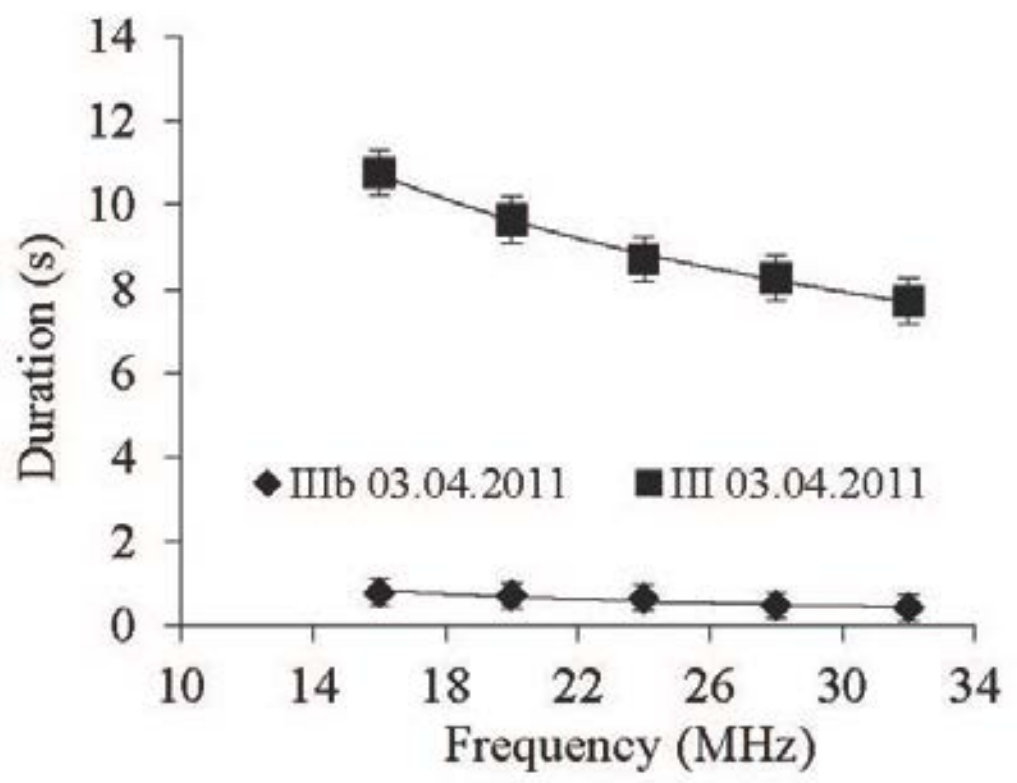

a)

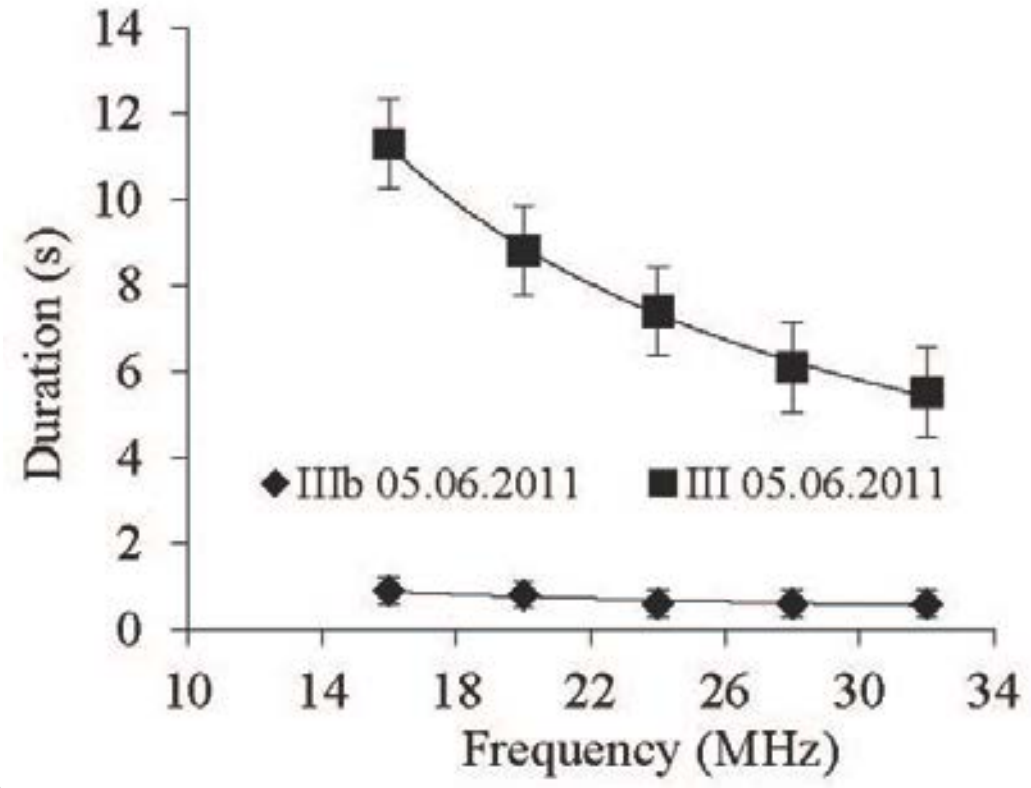

b)

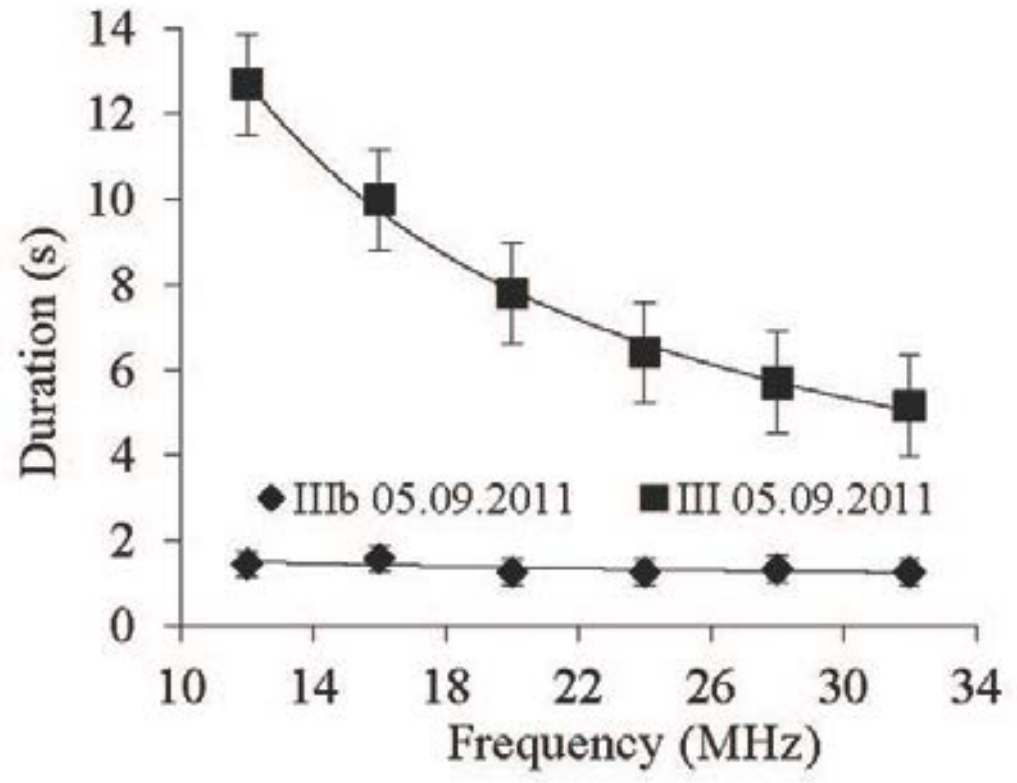

c)

Figure 9. Duration dependences for components of IIIb–III pairs on frequency for 3 April (a), 5 June (b), and 5 September (c) 2011.

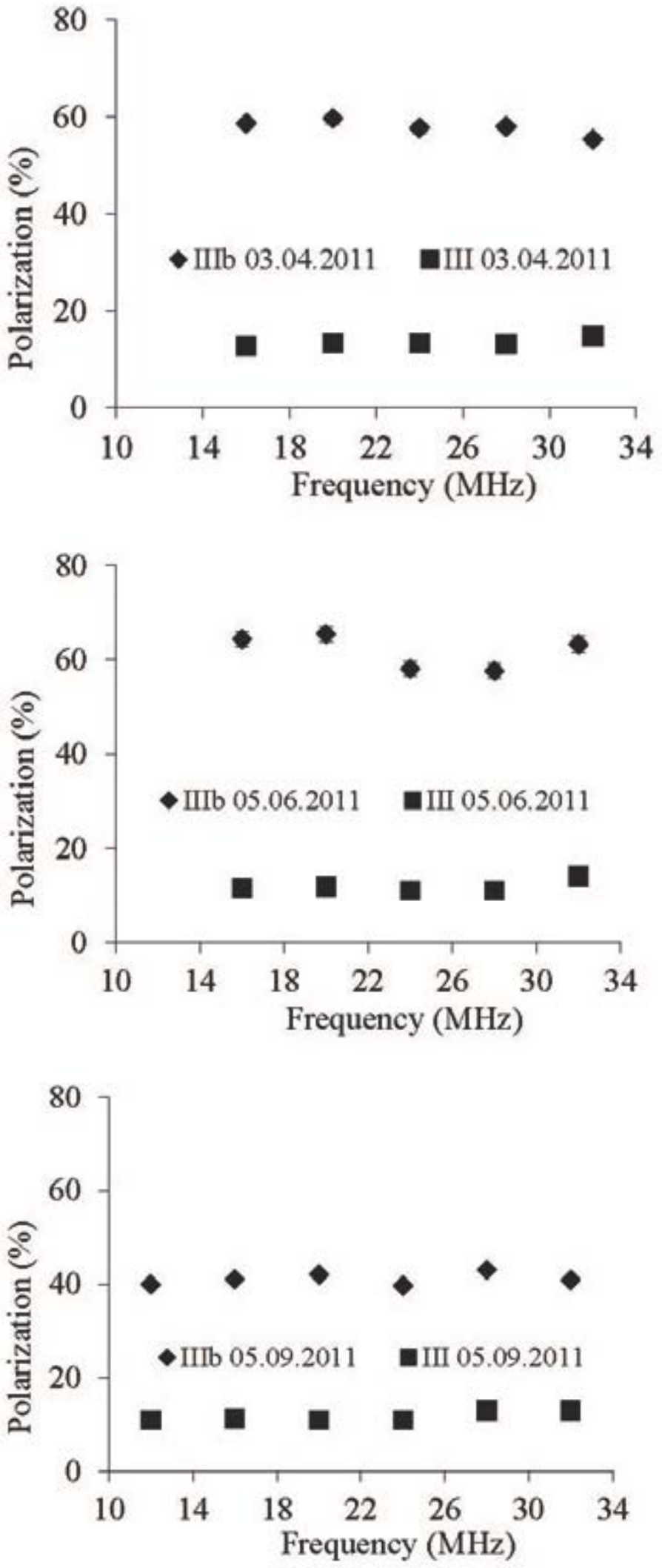

a)

b)

c)

Figure 10. Polarization dependences of components in IIIb–III pairs on frequency for 3 April (a), 5 June (b), and 5 September (c) 2011.

### 3.2 Frequency Drift Rate

As we have already seen, drift-rate dependences on frequency are linear (Equation 1) for both components of IIIb–III pairs. According to Melnik et al.(2011), coefficients A for powerful Type III bursts observed in 2002 had values from 0.08 to 0.12 that slightly differ from the values obtained for components IIIb– III pairs. As far as coefficients characterize inhomogeneity of coronal plasma $b = \left| dn / ndr \right|^{-1} = v_s / 2A$ ($v_s$ is the source velocity) (Melnik et al., 2011) then it can be said that at the location of generation of both components the inhomogeneities are approximately identical if the source velocities are the same. This is in favor of the harmonic character of IIIb–III components. Comparing coefficients A (from Equation 1) for different days, we conclude that the inhomogeneities in April and June were practically equal and are significantly smaller than that in September. One more argument supports harmonic connection of Type IIIb bursts and Type III bursts in IIIb–III pairs; this is the ratio of their drift rates. In the plasma mechanism of radio emission frequency drift rates for fundamental at frequency [ $f_I$ ] are defined by the equation

$$\frac{df_I}{dt} = \frac{df_{pe}}{dn}\frac{dn}{dr}\frac{dr}{dt} = \frac{f_{pe}}{2n}\frac{dn}{dr}v_s \qquad (3)$$

where $v_s$ is the source velocity and the density [n(r)] and the derivative [ $dn/dr$ ] are determined at the point of generation of radio emission at the local plasma frequency: $f_I = f_{pe}$. For the drift rate of the harmonic radio emission $f_{II} = 2f_{pe}$ the analogous equation

$$\frac{df_{II}}{dt} = \frac{d(2f_{pe})}{dn}\frac{dn}{dr}\frac{dr}{dt} = \frac{(2f_{pe})}{2n}\frac{dn}{dr}v_s \qquad (4)$$

is valid. Harmonic radio emission at frequency $f_{II} = 2f_{pe}$ means that it must happen at the location, where the plasma frequency is the same as in the previous case. So the density [n(r)] and the derivative [ $dn/dr$ ] are estimated at the same point. For this reason the ratio of drift rates at frequencies $f_{II} = 2f_{pe}$ and $f_I = f_{pe}$ equals

$$\frac{df_{II}}{dt} : \frac{df_I}{dt} = 2 \qquad (5)$$

The remarkable quality of this ratio is its independence from the model of the solar corona. As was mentioned earlier (Abranin, Bazelian, and Tsybko, 1984) these ratios were close to two for Type IIIb bursts and Type III bursts that have been observed respectively in the frequency intervals 6.25 – 12.5 MHz and

12.5 – 25 MHz. We derived that these drift rate ratios are equal to 1.9, 1.8, 1.6 for 3 April, 5 June, and 5 September, respectively. Taking into account errors in the measurement of drift rates, there is quite a good correspondence. The larger deviation of this ratio from the value two for 5 September is connected with a relatively small selection of data, only 23 pairs.

Comparing drift rates of Type IIIb bursts and Type III bursts at the same frequency in the case of constant source velocity gives a relation between coronal plasma inhomogeneities at heights that correspond to radio emission of fundamental and harmonic:

$$\frac{df_I}{dt} : \frac{df_{II}}{dt} = \frac{b_{II}}{b_I} \qquad (6)$$

Here $b_I = \left|dn/ndr\right|_I^{-1}$ and $b_{II} = \left|dn/ndr\right|_{II}^{-1}$ are sizes of inhomogenuities at the heights at which fundamental and harmonic are generated. The observational fact that Type IIIb drift rates are always larger than Type III burst drift rates in IIIb–III pairs denotes the natural condition that size inhomogeneity increases with distance from the Sun.

### 3.3 Duration

The essential distinction in the obtained durations of Type IIIb bursts and Type III bursts and their different frequency dependencies may give an indication of the various phenomena that define these durations. The durations of decameter Type III bursts are probably governed by the spatial sizes of the electron beams that are responsible for the generation of these bursts (Rutkevych and Melnik, 2012). In so far as in the plasma mechanism the radio emission originates at the local plasma frequency, then the frequency dependence of the duration and as a consequence the frequency dependence of the sizes of the electron beams shows how electron beam sizes change with height. This change is defined by the divergence θ of the magnetic field, along which the electron beams move. The numerical calculation (Rutkevych and Melnik, 2012) shows that at θ = 30° the power exponent [a] in the Equation 2 is equal to 0.6. At greater angles θ the electron beam size grows more quickly and accordingly the exponent [a] will be larger and vice versa. So we conclude that the angle θ was smaller than 30° on 3 April and was obviously larger than 30° on 5 June and 5 September.

Because Type IIIb bursts consist of stria-bursts, their durations are defined by stria durations. Decameter strias are very similar, especially in their dura- tions, to decameter spikes (Melnik et al., 2010). So it can be supposed that the durations of strias and spikes are governed by the same processes. As was shown by Melnik et al. (2014), durations of decameter spikes are determined by the collision time. For example, if the plasma temperature of low corona equals 1MK then the collision time is about one second at the heights 0.5 – 1 Rs and the spike durations are practically the same in the decameter band. We see from Figure 9 that Type IIIb burst durations really have the same values. If we suggest that Type IIIb burst durations are defined by collisions in plasma then the frequency dependence of their durations should be $\tau \propto n^{-1} \propto f^{-0.5}$ at constant temperature. Values of the obtained power [a] (Table 3) that differ from 0.5 can indicate that the plasma temperature changes with height.

Thus we see that durations of IIIb–III components can be interpreted in supposition that they are determined by the sizes of sources in one case and by collisions in another case. Why harmonic components (if this is correct) are defined by different causes remains unclear.

### 3.4 Polarization

As was mentioned above, the polarizations of Type IIIb bursts and Type III bursts in IIIb–III pairs were very different. This difference is a factor of three or four for 5 September and a factor of five or six for 3 April and 5 June. Such a difference can be understood in terms of plasma mechanism of radio emission if the components of IIIb–III pairs are the fundamentals and harmonics.

Radio emission in the decameter range escapes at the heights in the solar corona, where local plasma frequency $\omega_{pe} >> \omega_{Be}$ ($\omega_{pe} = \sqrt{4\pi e^2 n / m}$ is plasma frequency, $\omega_{Be} = eB / mc$ is the electron cyclotron frequency). So practically all plasma processes, which occur at these heights, are the same as in plasma without magnetic field (Melrose and McPhedran, 1991). There is one difference that is important for us. There are two electromagnetic waves, X-mode and O- mode, with different polarizations instead of one mode, electromagnetic waves, without polarization (Melrose and McPhedran, 1991). Dispersion relations for X-waves and O-waves are practically the same (Figure 11).

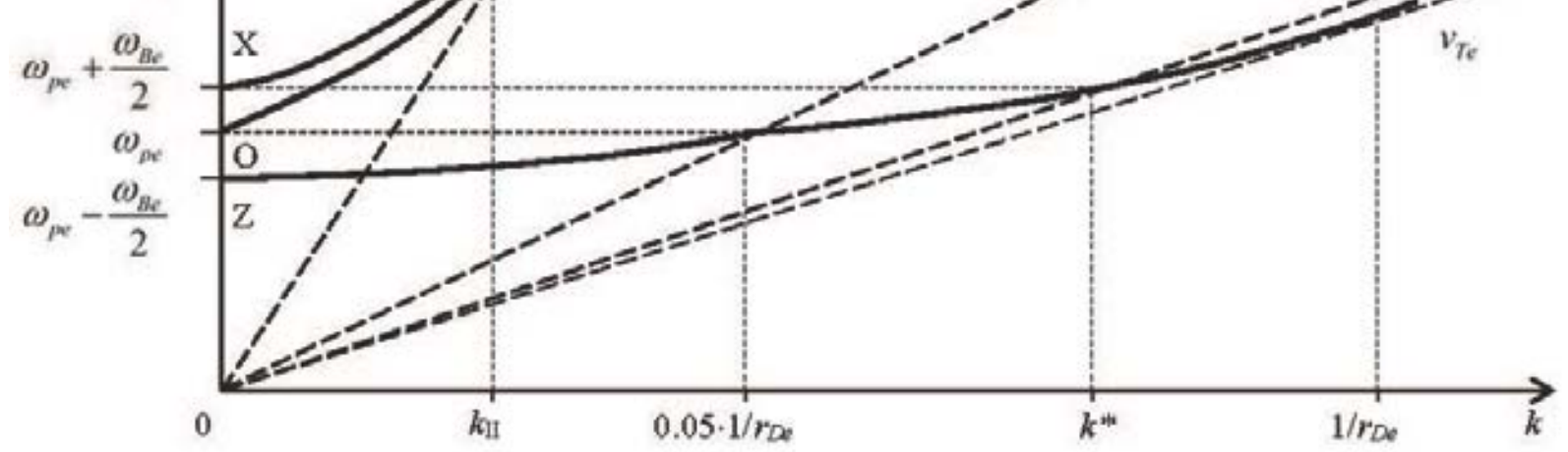

Figure 11. Dispersion curves for X-, O-, and Z-waves for plasma in a magnetic field.

Polarization of radio emission is defined by the difference of X- and O-waves (Melrose and McPhedran, 1991). The plasma mechanism is considered to be as such that the fundamental is generated due to conversion of Langmuir waves ($l$) into electromagnetic waves ($l$) in processes of scattering on ions ($i$) $l+i=t+i$ without changing frequency $\omega_l = \omega_I$ ($\omega_l$ is the frequency of Langmuir waves and $\omega_I$ is the frequency of electromagnetic waves) (Ginzburg and Zhelezniakov, 1958). Electron beams moving with velocity $v_s \approx 0.3c$ through coronal plasma generate Z-waves (which are Langmuir waves in plasma without magnetic fields) with the wave number $k \geq \omega_{pe} / v_s$. Mainly Z-waves are concentrated at minimum wave number $k_{\min} \approx \omega_{pe} / v_s$ (Mel'Nik, 1995, Mel'Nik, Lapshin, and Kontar, 1999). There is some amount of them at greater wave numbers up to $k_{\max} \approx \omega_{pe} / v_{Te} \approx 1 / r_{De}$ ($r_{De} = \sqrt{T_e / 4\pi e^2 n}$ is the Debye radius), but only a few of them. As seen from Figure 11 Z-waves with wave numbers from $k_{\min} \approx \omega_{pe} / v_s \approx 0.05 / r_{De}$ up to $k^*$ are transformed into O-waves and only waves with wave numbers $k > k^*$ are emerging into X-waves. So the number of O-waves will be significantly larger than that of X-waves and as a consequence the polarization of the fundamental will be considerable and its sign will coincide with O-waves. The wave number $k^*$ at a given plasma frequency is determined by the electron cyclotron frequency [$k^* = \sqrt{\omega_{Be}\omega_{pe} / 3v_{Te}^2}$], in other words by the magnetic field. So, e.g., at the magnetic field B = 1G the wave number $k^*$ is $0.18 / r_{De}$. In principle, knowing the spectral density of Z-waves and having measured the polarization of fundamental radio emission of Type IIIb bursts, it is possible to derive the magnetic field at the location of radio emission. Detailed consideration of this problem will be discussed in another article. Here we have shown that the fundamental has high polarization in the plasma mechanism of radio emission. The abovementioned fact that polarization maxima of Type IIIb bursts are achieved just at the beginning of the bursts, but not at their flux maximum, can be also understood. It was shown by Mel'nik, Lapshin, and Kontar (1999) that at the moment of first arrival of fast electrons in a certain place, the spectral energy density of Langmuir waves is centered at large phase velocities or small wave numbers. Langmuir waves with large wave numbers are practically absent. For this reason all Langmuir waves are transformed into O- waves and as a consequence polarization will be high. With time, the number of Langmuir waves with large wave numbers will be increasing. As a consequence the number of X-waves, into which these Langmuir waves are transformed, will also increase and thus burst polarization will decrease. Such behaviour of Type IIIb bursts is visible in Figure 6.

The second harmonic of radio emission appears in the processes of coalescence of two Z-waves (Mel'Nik and Kontar, 2003). In this respect the frequency of waves formed, either X-wave or O-wave, equals twice the local plasma frequency [$\omega_{II} = 2\omega_{pe}$]. The wave number of waves formed equals $k_{II} \approx \sqrt{3}\omega_{pe} / c$ and does not depend on the wave number of Z-waves. So numbers of formed X-waves and O-waves are defined by the probabilities of transformations of Z-waves into X- and O-waves (Zlotnik, 1981). Because these probabilities do not differed very much, then the number of X-waves and O-waves are not notably distinct, and as a result the polarization of the second harmonic is not high (Zlotnik, 1981). Thus analysis of polarizations of IIIb–III pairs show that the features of Type IIIb bursts and Type III bursts can be properly understood in terms of the plasma mechanism of radio emission in which the first components are the fundamental and the seconds are the harmonic.

Unfortunately, such important parameter as brightness temperature of Type IIIb and Type III bursts remains beyond the framework of our discussion but an adequate consideration this point can be apposite at measurements of IIIb– III pair sizes. At observations of discussed IIIb–III pairs their sizes were not measured. But lower measurements of single Type IIIb bursts and Type III bursts (Abranin et al., 1976, 1978) gave high brightness temperatures for both types of bursts that can be explained in the framework of plasma mechanism of radio emission (Mel'Nik and Kontar, 2003).

## 4. Conclusions

The properties of decameter IIIb–III pairs can be considered as such that Type IIIb bursts are the fundamental and Type III bursts are the second harmonic. In favor of this is the fact that frequency ratios of IIIb–III components are about two. As a consequence of harmonic character of components, the frequency drift ratios are also close to two. We consider as the very important evidence of harmonic structure of IIIb–III pairs the high polarizations of Type IIIb bursts and small polarizations of Type III bursts. Accompanying properties of these bursts, namely maximum polarization in advance of flux maximum for Type IIIb bursts and larger Type IIIb frequency drifts in comparison of Type III bursts frequency drifts, can also be explained.

The presence of Type IIIb fine structure in the form of stria-bursts and its absence for Type III bursts can be explained by small-scale inhomogeneities of coronal plasma (Loi, Cairns, and Li, 2014). At the same time there is an open question concerning the essential distinction of IIIb–III components durations. Moreover it is unclear what is the cause of the different physical phenomena that define these durations. Nevertheless in our opinion IIIb–III pairs are harmonic components because most of the observational indications are in favor of this idea.

**Acknowledgments** ∎

The work was partially financed in the frame of FP7 project SOLSPANET (FP7-PEOPLE-2010-IRSES-269299).